\documentstyle[11pt,newpasp,twoside]{article}
\def\edcomment#1{\iffalse\marginpar{\raggedright\sl#1\/}\else\relax\fi}
\reversemarginpar

\def\Msun{\hbox{$\thin M_{\odot}$}}
\def\Rsun{\hbox{$\thin R_{\odot}$}}
\def\thin{{\thinspace}}
\def\scr{\scriptstyle}

\def\tgs{{\thin \rlap{\raise 0.5ex\hbox{$\scr  {>}$}}{\lower 0.3ex\hbox{$\scr  {\sim}$}} \thin }}
\def\tls{{\thin \rlap{\raise 0.5ex\hbox{$\scr  {<}$}}{\lower 0.3ex\hbox{$\scr  {\sim}$}} \thin }}
\def\tll{{\raise 0.3ex\hbox{$\scr  {\thin \ll \thin }$}}}
\def\tgg{{\raise 0.3ex\hbox{$\scr  {\thin \gg \thin }$}}}
\def\tle{{\raise 0.3ex\hbox{$\scr  {\thin \le \thin }$}}}
\def\tge{{\raise 0.3ex\hbox{$\scr  {\thin \ge \thin }$}}}
\def\tl{{\raise 0.3ex\hbox{$\scr  {\thin < \thin }$}}}
\def\tg{{\raise 0.3ex\hbox{$\scr  {\thin > \thin }$}}}
\def\ts{{\raise 0.3ex\hbox{$\scr  {\thin \sim \thin }$}}}

\begin{document}
\title{Djehuty, a Code for Modeling Stars in Three Dimensions}
 \author{G. Baz{\'a}n, D. S. P. Dearborn, D. D. Dossa, P. P. Eggleton, 
A. Taylor, J. I. Castor, S. Murray, K. H. Cook, P. G. Eltgroth, R. M. Cavallo, 
S. Turcotte, S. C. Keller, and B. S. Pudliner}
\affil{Lawrence Livermore National Laboratory, Livermore CA, 94550}

\begin{abstract}
Current practice in stellar evolution is to employ one-dimen-
sional calculations that 
quantitatively apply only to a minority of the observed stars (single
non-rotating stars,
or well detached binaries).  Even in these systems, astrophysicists are
dependent on
approximations to handle complex three-dimensional processes like
convection.
Understanding the structure of binary stars, like those that lead to the
Type~Ia 
supernovae used to measure the expansion of the universe, are grossly
non-spherical
and await a 3D treatment.
 
To approach very large problems like multi-dimensional modeling of stars,
the Lawrence
Livermore National Laboratory has invested in massively parallel
computers and invested
even more in developing the algorithms to utilize them on complex
physics problems.
We have leveraged skills from across the lab to develop a 3D stellar
evolution code,
Djehuty (after the Egyptian god for writing and calculation) that
operates efficiently on
platforms with thousands of nodes, with the best available physical data
(opacities, EOS,
etc.). Djehuty has incorporated all basic physics for modeling stars
including an accurate
equation of state, radiation transport by diffusion, thermonuclear
reaction rates, and
hydrodynamics, and we have begun testing it in a number of applications.
\end{abstract} 
 
\section{Why do stellar modeling in three dimensions?}

The best determinations of the size, age and composition of the universe
are founded on
measurements of stars, combined with a physical understanding based on
one-dimensional (1D) computations. 1D calculations
quantitatively apply only to
a minority of the observed stars (single non-rotating stars, or well
detached binaries).
Even in these systems, astrophysicists are dependent on 1D approximations to
complex three-dimensional (3D) processes like convection.

Observations assure us that our best 1D approximation of steady state
convection is
flawed, requiring ad hoc `overshoot' corrections.  Even worse, there is
no real ability to
model the time-dependent convection that may be critical in modeling
helium flashes, Cepheid and Mira pulsations, or
the nucleosynthesis in deep mixing events that follow thermal pulses.
In addition, nearly
half of the visible mass of the universe is incorporated into
intrinsically 3D binary stars, many of which interact significantly at some stage of
evolution.
Understanding integral properties such as the chemical evolution, or the
detailed pre-explosion structure, of metric phenomena like type Ia
supernovae awaits development of a
true 3D stellar evolution code.

Developing a 3D code capable of modeling stars is exceedingly
challenging.  In spite of
the inherent difficulty, the importance of obtaining a better
understanding of stars has
led a small number of groups to begin such work.  Most current
calculations are
practically limited to of order 1 million zones, and cannot
realistically handle whole stars
(in 3D).  Envelope convection has been simulated by modeling modest
segments of
a
star (Clement 1993; Ludwig, Jordan, \&  Steffen 1994; Freytag, Ludwig,
\& Steffen
1996; Brummell, Hurlburt \& Toomre 1996, 1998; Stein \& Nordlund 1998; Skartlien, Stein, \& 
Nordlund 2000; Nordlund \& Stein 2001; Porter \& Woodward 2000).
Quite reasonably, these simulations often lack physical processes
pertinent to the core
(nuclear energy production), or radiation transport,  and the
gravitational potential is an
imposed condition.  While these are important starts toward
understanding convection
in stars, it is clear that some problems (e.g. solar rotation) yield
results that are
dependent on the size of the segment that can be simulated (Robinson \&
Chan 2001), necessitating whole star modeling.

With the advent of massively parallel computers and the development of
algorithms
able to operate in such a partitioned environment, it became possible to
consider
developing a code, Djehuty, capable of following meshes with sizes of
order $10^8$
cells and with a 
more complete suite of physics. Djehuty is  capable of modeling complete
stars in three dimensions.  It is based on the ALE (Arbitrary
Lagrange-Eulerian) hydrodynamic
method with a predictor-corrector Lagrange-Remap formalism, that is
second-order 
accurate in both time and space. It supports energy transport, via a
pair of diffusion
equations, one for matter conduction, and another for radiation. The
mesh, constructed
of multi-block logically rectangular hexahedrons, is allowed to be
non-orthogonal and can
conform to the developing instabilities, providing a more accurate
tracking.  Djehuty
permits domain decomposition for parallel operation, using MPI message
passing.

The physical data in Djehuty is identical to that used in the 1D
evolution code
that
provides initial structure models.  It has an equation of state 
based on the work of Eggleton, Faulkner \& Flannery (1973), 
as updated by Pols et al.
(1995; hereinafter PTEH),  that is very
accurate for stellar compositions
over a large range of masses ($\tg 0.5\Msun$). Both Planck and
Rosseland mean
opacities for radiation diffusion come from OPAL (Rogers \& Iglesias
1992) and
are
augmented by Alexander opacities (Alexander \& Ferguson 1994) for the lower
temperatures where
molecules are significant.  The nuclear reaction network includes
reactions for
hydrogen,
helium, and carbon burning, and the nuclear reaction rates come from
Caughlan \&
Fowler (1988).  Nuclear energy production can also be computed by a
tabular set
of
reactions. Neutrino losses are in tabular form from the work of Itoh et al.
(1989, 1992, with errata).

The gravity implementation is currently complete only for spherical
stars, but is adequate to begin scientific investigation of a host of
issues related to convection. In
the coming year, we will implement a generalized gravity solver to study
rapidly rotating
or binary stars.  We are still learning how to improve operational
efficiency, but have
begun a set of large scale, full physics, parallel runs aimed at our
first science objective,
the structure of convective cores in high mass main sequence stars. 

\section{The Physics in Djehuty}

\subsection{An Astrophysical Equation of State}

As an analytic approximation, the PTEH equation of state (EOS)
provides continuous
thermodynamic derivatives for hydrodynamic consistency. This EOS has
been modified
to reproduce tabulated OPAL values to an accuracy of
better than 1\% for the entire range of conditions expected in stars from
0.7 to $50\Msun$ (over
their whole evolution).  Stellar models as low as 0.5$\Msun$ can
be computed, with
differences of only about 2\% in their envelopes.

A comparison of the difference between the OPAL tables and the PTEH
EOS is shown
in Figure~1.  Over most of the temperature-density plane, the difference is
under a percent.
The upper left corner of the diagram is the region of complete
degeneracy (relativistic
and non-relativistic), important for white dwarfs, and the cores of red
giants.
 While
OPAL tables were not calculated there, the thermodynamics of this region
is well
represented by the Fermi relations with which the PTEH EOS is in
excellent
agreement. The upper right hand corner is included in the OPAL tables
but without the inclusion of relativistic effects.  Here the PTEH
code is presumably more accurate,
and the difference shown as green and red contours arise from a limitation
in OPAL
calculations.

Between the degenerate and non-degenerate regions there is a difficult
region where
Coulomb corrections to the pressure are very significant.  This shows as
a diagonal band
of blue contours where the OPAL and PTEH EOS's differ by more than
1\%.  It is a
region occupied by brown dwarfs, and very low mass stars ($\tls 0.5\Msun$), and
we will improve agreement there if we decide to investigate such objects
in the
future.

Successfully transferring a hydrostatic model to Djehuty (with minimal
transients)
requires that both codes utilize the same EOS.  The PTEH EOS has been
inserted into Djehuty and tested on a set of two and three dimensional
problems as well as our 3D
stellar model.  In a simple hydrodynamics problem, the accurate  EOS
call can be a
substantial component of the time.  The time cost of using this accurate
equation of
state was minimized by implementing a sub-cycling procedure as well as a
method
improving the convergence rate of the algorithm.  We have been able to
optimize
the
run time of the EOS without sacrificing the accuracy of
the results.  From
the first crude implementation of the accurate EOS, a
five-fold speed up
has been achieved.

%%%%%%%%%%%% FIGURE 1
\begin{figure}
%\plotone{bazan/Captures/fig1.ps}
\plotfiddle{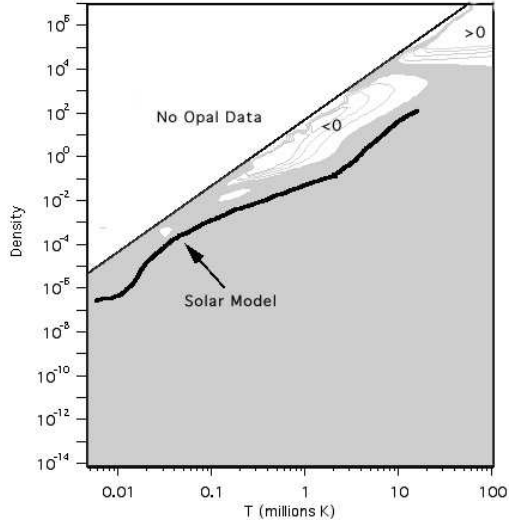}{7.5cm}{0}{45}{45}{-100}{0}
%\plotfiddle{EOS_fig1.ps}{7.5cm}{0}{45}{45}{-100}{0}
%\plotone{bazan/newfigs/EOS_fig1.ps}
%\vskip 6cm
\caption{The thick contours mark regions of no difference between
the analytic PTEH equation of state, and LLNL tabular values.  The grey region marks regions where they differ by $<1$\%. The diagonal line tracks the temperature density
profile of a solar model.}
\end{figure}
%%%%%%%%%%%% 

This EOS includes the formation of molecular hydrogen and ionization
states of low-Z elements.  While the EOS is able to operate on elements up to iron, it
is implemented to consider only $^1$H, $^3$He, $^4$He, $^{12}$C, $^{14}$N, $^{16}$O, $^{24}$Mg.

\subsection{Radiative Diffusion}

Energy transport via diffusion is an excellent approximation when the
cells of the mesh
are optically thick ($\tau \gg 1$ in a zone), a condition that is very well met
throughout a star's
interior. Djehuty  uses a pair flux-limited diffusion equations to
approximate energy
transport.  Included in the diffusion equation are sources and sinks
that link the radiation
equation to the hydrodynamic energy and momentum equations, as well as
nuclear
energy production, or neutrino losses. Djehuty operates in a
two-temperature (2T)
mode, in which the temperatures of the matter and of the radiation field
are integrated
separately, and the radiation is approximated as Planckian.  Again, this
is an excellent
approximation within stellar interiors, and will enable us to transition
to stellar
atmospheres.
These equations are:
\begin{equation}
{\rm radiation:}\ \ {\partial\phi\over\partial t}=\nabla.\thin({c\over 3\kappa_r\rho}\thin\nabla\phi) 
  + \kappa_p\rho c(aT^4-\phi)+E_\phi\ 
\end{equation}
\begin{equation}
{\rm matter:}\ \ \rho C_v{\partial T\over\partial t}=\nabla.\thin(D\thin\nabla T) 
  + \kappa_p\rho c(\phi-aT^4)+E_m\ \ .
\end{equation}
Here $a$ is the usual radiation constant, $\phi$ is the 
radiation energy density per unit volume (which would be $aT^4$ in complete
equilibrium), and $E_{\phi}, E_m$ are contributions to the radiation
and matter energies from compression, nuclear reactions etc. $D$ is the
heat transport coefficient by conduction alone.

Within a star, the matter and radiation are so strongly coupled that the
matter
and
radiation temperatures are essentially identical.  This strong coupling
is achieved because of the
large value for the Planck opacity, $\kappa_p$.  
%Currently, a Kramer's
%approximation is
%used for $\kappa_p$,
%%and is sufficiently large to ensure equality of the matter and radiation
%temperatures.
%More accurate calculations of $\kappa_p$ are being developed from OPAL data, and
%will be 
%implemented in the near future.  
A table of Planck opacities has been constructed from data from the
Opacity Project (Seaton et al. 1994) at high temperatures and 
of Alexander \& Ferguson (1994) at low temperatures.
Accurate Planck opacities will be
important near
the surface where the diffusion approximation becomes poor.
 
The transfer of radiation is governed by the Rosseland mean opacity,
$\kappa_r$.
Currently
Djehuty reads $\kappa_r$ from a set of astrophysical opacity tables that were
formed using OPAL (Livermore) opacities at high temperatures ($\tg 10,000\thin$K)
and Alexander opacities at lower
temperatures where molecules are important (Alexander \& Ferguson 1994; 
Rogers \& Iglesias 
1992).  This is the same opacity set used by the 1D evolution code.
These astrophysical
opacities will be the only option actually used in the stellar
calculations done by
Djehuty, but an analytic Kramer's form for the Rosseland mean opacity is
available to
facilitate comparisons with other codes.
 
The radiation package of Djehuty has been given some simple tests to
ensure accuracy.
The first was a simple Marshak wave problem.  A temperature source was
applied to one
end of a tube.  With hydrodynamic motions turned off, and a constant
opacity applied, the results can be compared directly with theory.
Excellent agreement was found,
providing direct evidence that the radiation diffusion package in
Djehuty functions well.
 
\subsection{Thermonuclear Rates and Neutrino Losses}

Elements normally carried directly by the 1D evolution code were
selected to allow
accurate tracking of principal energy generation reactions for hydrogen,
helium and carbon burning over the whole range of stellar masses.  This list of elements
includes $^1$H, $^3$He, $^4$He, $^{12}$C, $^{14}$N, $^{16}$O, $^{24}$Mg.  
These elements are passed to Djehuty as part of
the mapping process, and form a standard 7-element set.

The 7-element set is an excellent approximation for circumstances of
stable hydrogen burning in most stars.  It includes the slower hydrogen
burning reactions of the PP1 and
PP2 chains as well as the CNO cycles. Over evolutionary time-scales,
many reactions in
the PP chain and CNO Cycles come to equilibrium, and do not need to be
specifically
included.

The rate equations for the 7-element set are identical to those used in
the 1D code and
Djehuty integrates those rates with the abundances to determine local
energy production
and transmutation rates. In this mode, energy production computed by
Djehuty and the
1D code are well matched.  The rate equations used by Djehuty and the
1D code include
not just hydrogen burning, but the most important energy production
rates from helium burning and carbon burning stages. We have also included
strong and weak Coulomb screening.

Both Djehuty and the 1D code have been written to permit easy expansion
of the
element set. Djehuty can also operate with an 18-element set suitable for
situations
where the hot CNO cycle operates.  The program is now structured such
that it is trivial
to add new rates connecting existing elements.  It is also relatively
easy to increase the
element sets when circumstances require.

Operation of the new version is independent of the order in which the
elements are
specified in the input deck, and the solution of all rate equations is
implicit.  In tests, the
hydrogen mass fraction can be changed by orders of magnitude in a single
step.
As part of these changes the EOS was modified to handle an arbitrary element set.
 
% \begin{table}
% \caption{Standard 7-element suite (order does not matter)}
% \begin{tabular}{cccc}
 % species &  clyde & atwt & zbar  \\
% \hline
 % $^1$H    &  1001  & 1.0  & 1.0   \\
 % $^3$He   &  2003  & 3.0  & 2.0   \\
 % $^4$He   &  2004  & 4.0  & 2.0   \\
 % $^{12}$C &  6012  & 12.0 & 6.0   \\
 % $^{14}$N &  7014  & 14.0 & 7.0   \\
 % $^{16}$O &  8016  & 16.0 & 8.0   \\
 % $^{24}$Mg&  12024 & 24.0 & 12.0  \\
% \end{tabular}
% \end{table}
 % 
% \begin{table}
% \caption{Standard 18-element suite (order does not matter)}
% \begin{tabular}{cccc}
 % species &  clyde & atwt & zbar  \\
% \hline
 % $^1$H   &  1001    &  1.0  & 1.0  \\
 % $^3$He  &  2003    &  3.0  & 2.0  \\
 % $^4$He  &  2004    &  4.0  & 2.0  \\
 % $^{12}$C &  6012    &  12.0 & 6.0  \\
 % $^{13}$C &  6013    &  13.0 & 6.0  \\
 % $^{13}$N &  7013    &  13.0 & 7.0  \\
 % $^{14}$N &  7014    &  14.0 & 7.0  \\
 % $^{15}$N &  7015    &  15.0 & 7.0  \\
 % $^{15}$O &  8015    &  15.0 & 8.0  \\
 % $^{16}$O &  8016    &  16.0 & 8.0  \\
 % $^{17}$O &  8017    &  17.0 & 8.0  \\
 % $^{18}$O &  8018    &  18.0 & 8.0  \\
 % $^{17}$F &  9017    &  17.0 & 9.0  \\
 % $^{18}$F &  9018    &  18.0 & 9.0  \\
 % $^{19}$F &  9019    &  19.0 & 9.0  \\
 % $^{22}$Ne &  10022   &  22.0 & 10.0  \\
 % $^{24}$Mg &  12024   &  24.0 & 12.0  \\
 % $^{32}$S  &  16032   &  32.0 & 16.0  \\
% \end{tabular}
% \end{table} 

In both element sets, the proton-proton chain is handled as follows:
the proton capture on deuterium, p(p,$\beta\,\nu$)D(p,$\gamma$)$^3$He, is assumed to
be followed instantly by $^3$He($^3$He,2p)$^4$He and $^3$He($^4$He,$\gamma$)$^7$Be(p,$^4$He)$^4$He.
 
In the 7-element set, only the slower rates are included in the CNO
cycle. The beta decay and proton capture on $^{13}$C, 
$^{12}$C(p,$\gamma$)$^{13}$N($\beta$,$\nu$)$^{13}$C(p,$\gamma$)$^{13}$N,
are assumed to be instantaneous.
The beta decay and proton capture on $^{15}$N, 
$^{14}$N(p,$\gamma$)$^{15}$O($\beta$,$\nu$)$^{15}$N(p,$\alpha$)$^{12}$C,
are also assumed to be instantaneous, as are
the beta decay and proton capture on $^{17}$O,
$^{16}$O(p,$\gamma$)$^{17}$F($\beta$,$\nu$)$^{17}$O(p,$\alpha$)$^{14}$N.

\def\nC{{\rm C}}
\def\np{{\rm p}}
\def\nN{{\rm N}}
\def\nF{{\rm F}}
\def\nO{{\rm O}}
All CNO rates are included in the 17-element suite making it suitable
for the hot CNO
cycle, including leakage into $^{19}$F. They are:
$$^{12}\nC(\np,\gamma)^{13}\nN,\ \     ^{13}\nN(\beta,\nu)^{13}\nC,\ \    ^{13}\nC(\np,\gamma)^{14}\nN,
\ \ ^{14}\nN(\np,\gamma)^{15}\nO,\ \     ^{15}\nO(\beta,\nu)^{15}\nN,$$
\vskip -0.4truein
$$\ \    ^{15}\nN(\np,\alpha)^{12}\nC,
\ \ ^{15}\nN(\np,\gamma)^{16}\nO,\ \     ^{16}\nO(\np,\gamma)^{17}\nF,\ \   ^{17}\nF(\beta,\nu)^{17}\nO,
\ \ ^{17}\nO(\np,\alpha)^{14}\nN,$$
\vskip -0.4truein
$$\ \     ^{17}\nO(\np,\gamma)^{18}\nF,\ \   ^{18}\nF(\beta,\nu)^{18}\nO,
\ \ ^{18}\nO(\np,\alpha)^{15}\nN, \ \ {\rm and}\ \    ^{18}\nO(\np,\gamma)^{19}\nF.$$

The following helium burning reactions are also included:
 $^4$He(2$\alpha$,$\gamma$)$^{12}$C, $^{12}$C($\alpha$,$\gamma$)$^{16}$O,     
$^{14}$N($\alpha$,$\gamma$)$^{18}$O, and 
$^{18}$O($\alpha$,$\gamma$)$^{22}$Ne.    
In the 7-element set, the last reaction is assumed to
happen instantaneously, and the mass fraction
change is placed with all other heavy elements in
$^{24}$Mg.

Finally, the following reactions are included for beginning advanced
stages of massive star evolution:
$^{12}$C($^{12}$C,$\gamma$)$^{24}$Mg and $^{16}$O($^{16}$O,$\gamma$)$^{32}$S.

As a result, the 7-element set uses 12 reaction rates and the 18-element
set evaluates 23 rates.  These rates are all evaluated from equations following the
form of Harris et al. (1983) and returned as reactions per mole per
second.  
How the reactions are used is defined by three numbers returned for each
rate that specify the number of input particles, of output particles,
and the number of particles in reaction.
%How the reactions are used is defined in three integer arrays that are
%returned with the rates, {\tt num}, {\tt in}, and {\tt iout}.
%
%The {\tt num} array stores three numbers for each rate:
%{\tt num(irate,1)} stores the number of  input particles,
%{\tt num(irate,2)} stores the number of output particles, and
%{\tt num(irate,3)} stores the number of particles in reaction.
%
This permits the code to include appropriate factors for true multi-body
rates like the
triple-alpha rate. It also allows two-body rates, like $^{12}$C(p,$\gamma$)$^{13}$N, to
be correctly evaluated
while reducing the number of protons by 2 in the approximation where
the reactions
$^{13}$N($\beta$,$\nu$)$^{13}$C(p,$\gamma$)$^{14}$N are assumed instantaneous.
 
%In order to specify the elements in and out in the arrays {\tt in} and {\tt iout},
%the program must
%know the order with which the elements were specified.  This is tested
%at the beginning of the subroutine `burnit'.

Each rate also includes the net energy produced from the rate (less any
neutrino  energy).  This information is used along with the rates to derive the
energy production rate in ergs per gram per second.  A table for neutrino losses 
(Itoh et al. 1989, 1992) from
Compton, bremsstrahlung, and pair production processes has been included, and is
initialized along with the basic data for the EOS routine.
%(the call to
%``const'' for initialization occurs in
%``main'').  
The temperature and density are then used to interpolate in
this table to provide a neutrino loss rate.

Finally, Djehuty can be quickly transformed by providing the rates in
tabular form.  We
have generated a table using rates associated with the 7-element set and shown good
correspondence in energy production and transmutation rates. The tabular
method can be implemented without recompiling Djehuty, and lends itself to quick
tests.  In the future we anticipate using it to simulate chemical rates for modeling
giant planet atmospheres.

\subsection{Gravity}

At present we have implemented a simple algorithm to compute the gravity
of a
spherical object.  Between hydrodynamic cycles, Djehuty integrates over
all cells of the
mesh to determine a mass-radius relation.  In parallel operation, this
requires
sharing a
relatively small amount of information among the processors to construct
the relation
for the entire star.

Once the spherical mass distribution has been found, the gravitational
acceleration can
be simply calculated at each node.  This method is limited to spherical
objects, but its
implementation identified much of the process that must occur for a more
general 
treatment to be implemented.

As an extension of this self-consistent spherical gravity, it is
possible to add a point
mass outside of the problem, and study tidal modification of the model.
This approximation is appropriate only when the star is sufficiently
centrally condensed that
the portions of the star that are distorted are of modest mass.

\subsection{Hydrodynamics}

Djehuty uses an ALE (Arbitrary Lagrange-Eulerian) hydrodynamic method,
in which
an
explicit hydrodynamic step is followed by an advection step.  In stable
regions, or regions
of large-scale coherent motion where the Lagrangian mesh is modestly
deformed, no
advection is necessary, and the code is essentially Lagrangian.  In
regions where shear
develops, it is necessary relax the mesh, and permit material to move
between cells.

To promote hydrodynamic accuracy, and time centering for other physical
processes, the
hydrodynamic step is split in half.  The first sub-step is a simple
explicit step.  This is
followed by a second sub-step in which information from both the
original position, and
information from the half (sub) step position, are used in a
predictor-corrector
formalism.
This allows a time-centered inclusion of radiation transport and energy
production with a
hydrodynamics step that is second-order accurate in both time and space.
In Djehuty, vector quantities like position and velocity are zone-centered, 
while scalar quantities
(density, temperature ...) are cell-centered. By splitting the
hydrodynamic step, the node
and cell quantities have the same time.

\section{Model Generation}

\subsection{The Initial Models}

We developed a 1D stellar evolution code for use as a platform to test
physics, and to
provide structure information to Djehuty for constructing 3D models.
Djehuty  reads the
1D stellar evolution models directly, and converts the model into its
own variable set.
The evolution code follows the work of Eggleton (1972) and has a fully implicit
algorithm that
includes mesh motion during a time step.  This speeds the calculation
and improves its
accuracy, but results in an unusual centering of the thermodynamic
variables.  To
minimize any transients, software was also written to adjust (and relax)
the variables to
values appropriate for cell-centered equations used by Djehuty.

As an example the 1D code is capable of following a $4\Msun$ model
from a pre-main-sequence configuration through main sequence, giant,
and thermally pulsing stages to becoming a white dwarf (Figure~2).  At
any stage of evolution, a restart file can be written that is usable
by either Djehuty or the 1D code itself.
 
%%%%%%%%%%%%%% FIGURE 2
\begin{figure}
%\plotone{\arg{file}}
\plotfiddle{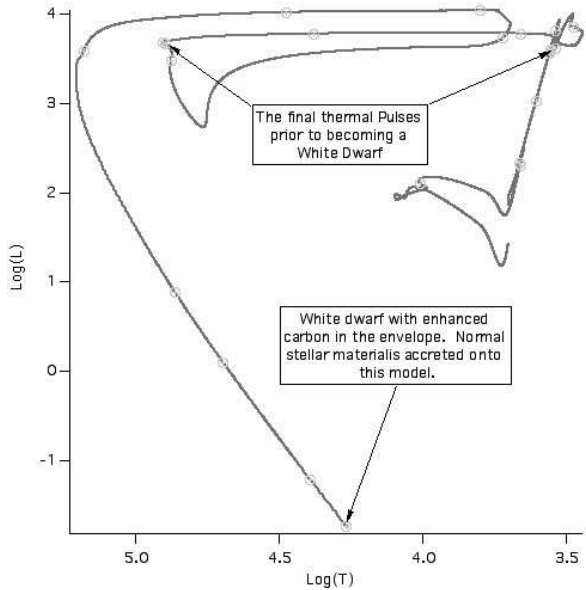}{7.5cm}{0}{45}{45}{-140}{0}
%\plotfiddle{HR_fig2.ps}{7.5cm}{0}{45}{45}{-140}{0}
%\plotone{bazan/newfigs/HR_fig2.ps}
%\vskip 4cm
\caption{An HR diagram shows the temperature luminosity evolution of a $4\Msun$
star from the pre-main-sequence to a point where its remnant core has become a white
dwarf. The model positions are shown by open circles.}
\end{figure}
%%%%%%%%%%%%%

As a first test of Djehuty, a pre-main-sequence model of a star
approaching the main
sequence was tested.  At this point, nuclear reactions are negligible
($E_{nuc}/E_{grav}\tl 0.005$),
the composition homogeneous, and the star is almost entirely stable.
This model was
selected so as to test the accuracy of our scaling and mapping algorithms
as well as to test various 3D mesh structures.

A  second model that was selected for an initial test case was of a star
that has just reached the main sequence. It has a convective core of about $0.4\Rsun$
($2.8\times 10^{10} \hbox{\rm cm}$),
driven by the strong temperature sensitivity of CNO nuclear
burning.  In this model (Figure~3),
carbon has been converted into nitrogen, the core $^3$He has been
destroyed, and the
central hydrogen mass fraction has fallen from 0.7 to 0.676.  This model
has been used
to test the nuclear energy generation rates as implemented in Djehuty.
 
%%%%%%%%%%%%%% FIGURE 3
\begin{figure}
\plotfiddle{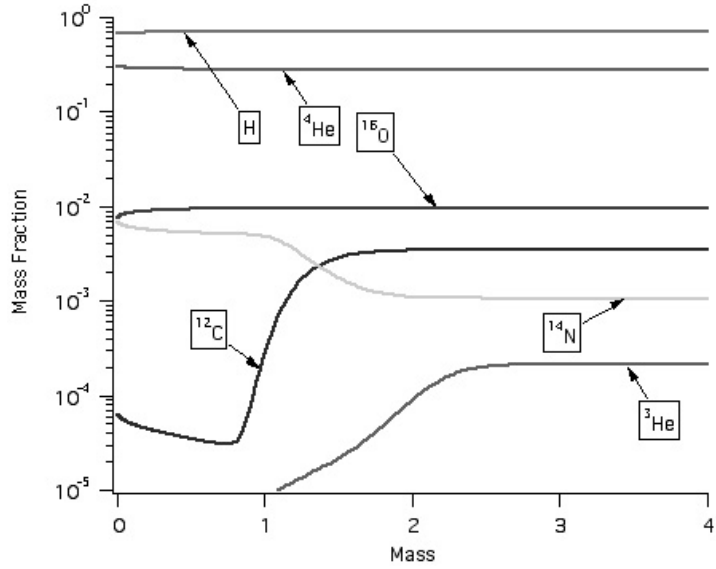}{7.5cm}{0}{60}{60}{-140}{0} 
%\plotfiddle{Composition_fig3.ps}{7.5cm}{0}{60}{60}{-140}{0} 
%\vskip 6cm
\caption{The composition structure of an early main sequence star of $4\Msun$.}
\end{figure}
%%%%%%%%%%%%%

\subsection{Mesh Structure}

In order to reduce the problem of tiny zones (and time steps), inherent
in a mesh using
spherical coordinates, we are utilizing a technique in which the sphere
is constructed from
7 logical blocks.  There is a central block, and 6 adjacent blocks, one
attached to each
face of the central block.  The mesh in each of the outer blocks is
morphed to form a
spherical tile.  The central block can remain as a cube or be
morphologically distorted
from near rectangular cells at the center to a sphere at its surface.
Preliminary numerical
experiments suggest that allowing the central block to retain most of
its cubic
shape
results in a better-behaved structure, and smaller deviations from
hydrostatic equilibrium
(particularly at the corner nodes of the central cube).

As an example (Figure~4), we show a sample logical structure in which the central
block (0) is
constructed from a mesh with 50x50x50 cells (51x51x51 nodes).  There are
then 6
additional blocks (1-6), having 50x50x100 cells.  In each case, the axis
oriented in the
radial direction has 100 zones (101 nodes).  Blocks 1, 2, and 3, attach
to the faces 
on which the logical variables i, j, and k are 1, while blocks 4, 5, and
6 attach to the
faces where i, j, and k have their maximum values.

%%%%%%%%%%%%%%% FIGURE 3
\begin{figure}
%\plotone{arg{file}}
\plotfiddle{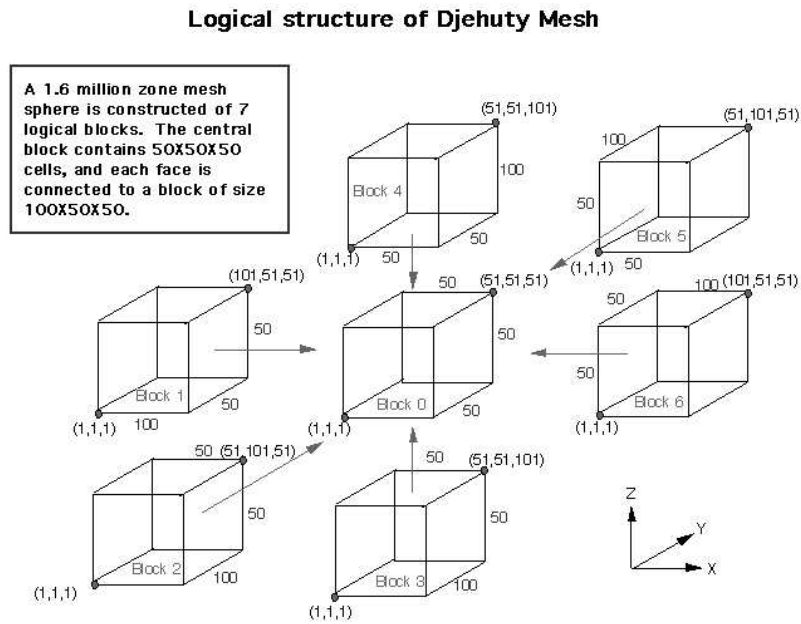}{8cm}{0}{70}{70}{-150}{0}
%\plotfiddle{BlocksBW_fig4.ps}{8cm}{0}{70}{70}{-150}{0}
%\vskip 4cm
\caption{Logical structure of Djehuty mesh.}
\end{figure}
%%%%%%%%%%%%%%

The logical (surface) blocks that attach to the central cube must (at
present) have the same mesh size as the face to which they adjoin, but
the radial component may have any length. These surface blocks are
transitioned from a cubic shape to spherical segments (Figure~5).  An
initial mesh is generated with a standard size. It can then be
tailored to a particular stellar model (radius and radial zoning
structure).  The scaling must not just match the radius of the star,
but have a radial spacing to resolve important gradients.  The radial
structure of the 1D mesh is used as an initial guide for scaling the
3D mesh, but the user has an ability to bias the mesh through a
parameter.

This additional flexibility is provided  by producing a scale factor
that is parameterized as
`grid\_scale\_alpha'.  
The scale factor, $S_r$, was just defined as
\begin{equation}
  S_r = r^\alpha
\end{equation}
where $r$ is the fractional radius of a node in the mesh.  Setting
grid\_scale\_alpha to a value of 1 results in a direct mapping of the
radial structure from the 1D model to the 3D mesh.  A value of
grid\_scale\_alpha greater than 1 concentrates the mesh in the core,
and makes the envelope sparser.  Alternatively a value less than 1
enlarges the core and refines the envelope (Figure~6).

%%%%%%%%%%%%%%% FIGURE 5
\begin{figure}
\plotfiddle{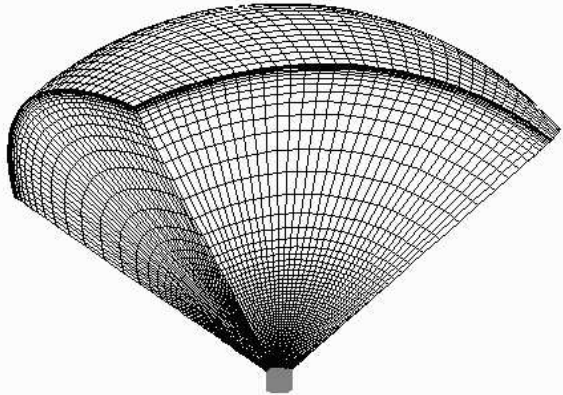}{5.5cm}{0}{50}{50}{-110}{0}
%\plotfiddle{WedgeBW_fig6.ps}{5.5cm}{0}{50}{50}{-110}{0}
\caption{The structure of a representative mesh with 590K zones}
\end{figure}
%%%%%%%%%%%%%%
%%%%%%%%%%%%%%% FIGURE 6
\begin{figure}
\plotfiddle{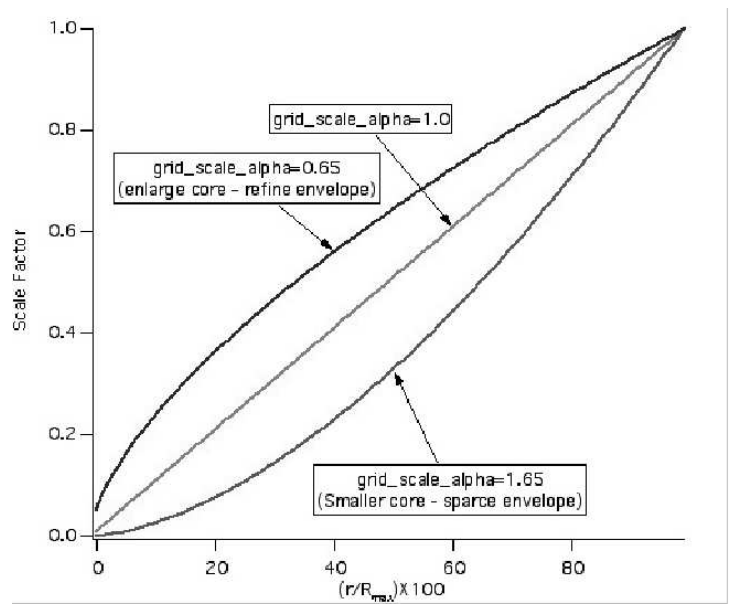}{7cm}{0}{50}{50}{-170}{-70}
%\plotfiddle{fig4.eps}{7cm}{0}{50}{50}{-170}{-70}
\caption{In a coarse 3D mesh, 3 to 4 radial zones from the 1D model may
contribute to the properties of a cell.}
\end{figure}
%%%%%%%%%%%%%%
%%%%%%%%%%%%%%% FIGURE 8
\begin{figure}[h!]
%\begin{figure}
%\plotone{newfigs/density_fig8.ps}
%\plotone{density_fig8.eps}
\plotfiddle{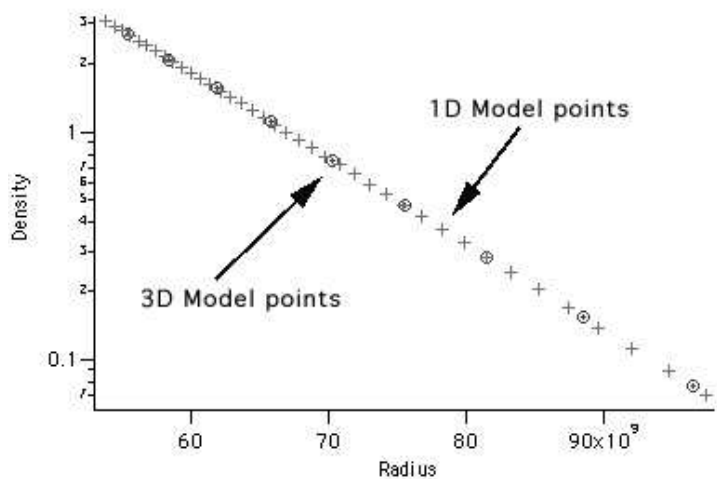}{7cm}{0}{70}{70}{-150}{0}
\caption{Comparison of 1D model to Djehuty mesh.}
\end{figure}
%%%%%%%%%%%%%%

Once the mesh has been scaled, the 1D model is mapped into it.  The
cells of the small meshes that we have used for our initial testing
are of such a size that 3 or 4 of the 1D radial zones contribute to
the properties of the cells.  Furthermore, the cells near the core
will not (in general) have a radial orientation. A simple mapping of
zone-centered values led to a cumulative error in total stellar mass
that was unacceptably large.  This mapping problem was solved by
subdividing our 3D cells and integrating the 1D model across the
smaller volume elements (Figure~7).  The segment masses and volumes
were then summed to determine the average cell density.  Other
properties were then mass-averaged among the segments.

%%%%%%%%%%%%%%% FIGURE 9
%\begin{figure}
%\plottwo{\arg{file1}}{\arg{file2}}
%\plotfiddle{bazan/Captures/fig6.eps}{7cm}{0}{50}{50}{-170}{-100}
%\vskip 18cm
%\caption{Cell definition}
%\end{figure}
%%%%%%%%%%%%%%

%%%%%%%%%%%%%%% FIGURE 10
\begin{figure}[h!]
%\plottwo{\arg{file1}}{\arg{file2}}
\plotfiddle{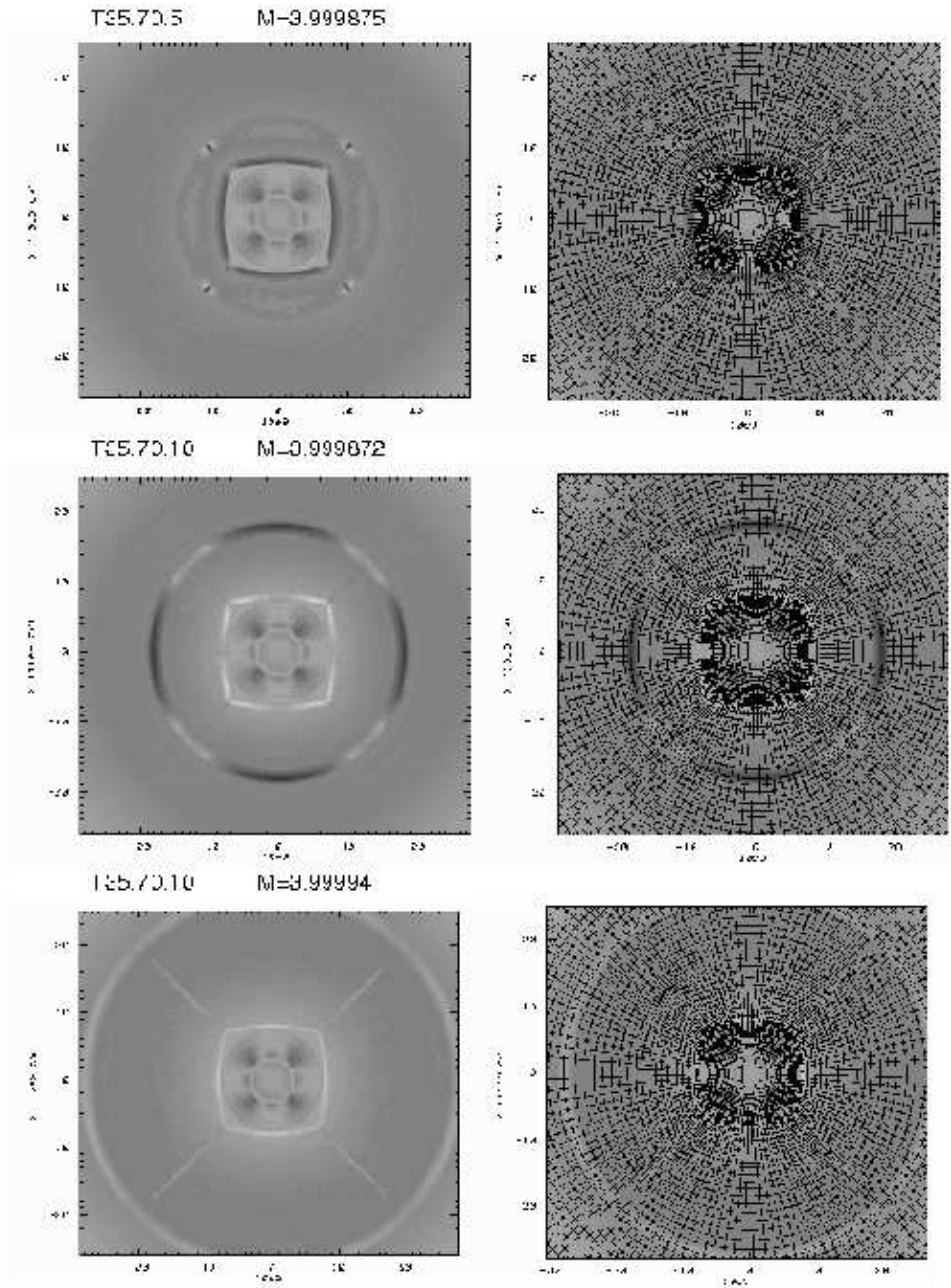}{18.3cm}{0}{82}{82}{-250}{-90}
%\plotfiddle{fig8.eps}{18.3cm}{0}{82}{82}{-250}{-90}
%\vskip 4cm
\caption{Mesh structures. The initial velocity transients are minimized when the transition region is smoother.}
\end{figure}
%%%%%%%%%%%%%%

%%%%%%%%%%%%%%% FIGURE 11
%\begin{figure}
%\plottwo{\arg{file1}}{\arg{file2}}
%\plotfiddle{bazan/Captures/fig9.eps}{18cm}{0}{75}{75}{-230}{-20}
%\vskip 4cm
%\caption{A second order pre-conditioning scheme reduces mapping error to hydrostatic equilibrium, as dose the use of the more accurate EOS 550.}
%\end{figure}
%%%%%%%%%%%%%%

%%%%%%%%%%%%%%% FIGURE 12
%\begin{figure}
%\plottwo{\arg{file1}}{\arg{file2}}
%\plotfiddle{bazan/Captures/fig10.eps}{18cm}{0}{75}{75}{-230}{-40}
%\caption{Shows the refined core of 1D mesh (600 zone model). Changing the mesh spacing coefficents in the 1D code reduces the size of the central zone and minimizes core transients.}
%\end{figure}
%%%%%%%%%%%%%%
\section{Model Settling}

Figure~8 gives an example of the initial velocity perturbations that
result from the mapping of a $4\Msun$ model to a 3D mesh.  The model
is generated with a fixed boundary condition, and run for 5 cycles to
an elapsed time of 0.074416 seconds.  The grey scale ranges from 0 to
500 cm/s.

\section{Conclusion}

Djehuty is now an operational code, and papers are in preparation on
stellar simulations that have been made with it (see Figure~3 of
Eggleton et al. in these proceedings). Highlights of these
calculations include the first 3D simulation of a type II
supernova. The nickel produced in the silicon-burning region of these
stars is an important energy source in the observed light curves.  In
SN1987a, it appeared much sooner that expected from a 1D model.  Our
3D calculation shows the development of an instability causing
tendrils of nickel to extrude into surrounding material, where it
appears early, as seen by observers.  We have also done a first
simulation of a nova.  Novae are explosive events that occur in
binary-star systems where hydrogen is being torn from a normal star
and deposited on to a white dwarf companion.  In our first 3D
calculation, the hydrogen ignition propagated from a number of
ignition points, and created a shock that accelerated the surface to
near 3000 km/s. When the shock emerges at the surface, it is heated to
temperatures in excess of 1 Kev, creating an X-ray pulse.  In this
first calculation, the white dwarf mass is a bit smaller that those
thought appropriate for most novae, and the hydrogen accretion was
symmetric over the surface of the star.  A more realistic hydrogen
distribution in which the material is settled on to the star as a high
velocity stream will be a future calculation. Simulating stable stars,
like our $4\Msun$ main sequence model, has been remarkably
challenging, but with new boundary conditions the numerical stability
problems that precludes long runs seem to have been resolved.

\acknowledgments
This work was performed under the auspices of the U.S.
Department of Energy, National Nuclear Security Administration by the
University of California,
Lawrence Livermore National Laboratory under contract No. W-7405-Eng-48.

%\plotone{\arg{file}}
%\plottwo{\arg{file}}{\arg{file}}
%\plotfiddle{\arg{file}}{\arg{vsize}}{\arg{rot}}{\arg{hsf}}{\arg{vsf}}{\arg{htrans}}{\arg{vtrans}}

%\vfill\eject
\end{document}